\documentclass{durville}
\usepackage[dvips]{graphicx}

\title{APPROACH OF THE CONSTITUTIVE MATERIAL BEHAVIOUR OF TEXTILE COMPOSITES THROUGH SIMULATION}

\author{Damien Durville}

\heading{Damien Durville}

\address{LMSSMat\\
Ecole Centrale Paris / CNRS UMR 8579
Grande Voie des Vignes \\
92295 Ch\^atenay-Malabry Cdex - France\\
e-mail: damien.durville@ecp.fr, web page: http://www.mssmat.ecp.fr}


\abstract{A complete approach for the determination of the complex constitutive behaviour of textile
  composites through finite element simulation is presented in this paper. In this work, simulations
  of different loading cases are carried out on small samples of textile composites, taking into
  account all individual fibers of the fabric and interactions taking place between them. The most
  delicate issue is related to the modelling of contact-friction interactions between fibers, and
  the solution of the nonlinear problem that follows up by the means of robust algorithms. The
  efficiency of these algorithms, and the increase of capacities of high performance computing,
  allow to simulate samples made of several hundreds of interlocked fibers.}

\begin{document}

\section{INTRODUCTION}

Textile composites are constituted by components arranged at different levels : fibers are spun
together to form yarns that are then interlocked and coated with an elastic matrix. The prediction
of the mechanical properties of such structures requires then to be able to represent the behaviour
of each component, and above all, to be able to take into account interaction mechanisms taking
place between them. However, relative motions between internal fibers are very complex, and
classical techniques, such as homogeneization, most of the time reveals helpless to account for
these phenomena.

With the increase of computing capacities, the simulation becomes an alternative way to explore and
identify the behaviour of textile materials. To carry out such a simulation, one has to model the
behaviour of internal components and their interactions. At a first level, one can consider yarns
making up the fabrics by the means of adapted beam model \cite{Durv03,Boisse05}. But the behaviour
of individual textile yarns are also very complex, especially in their transverse directions, and
for instance, the shape of their cross-section may vary largely depending on the position in the
weaving, and the tension in the yarns.

It is now possible to consider a large number of interacting fibers, and to go down one level
deeper, in order to take into account the behaviour of individual fibers constituting the yarns.
Based on our approach to model contact-friction interactions within a large collection of fibers
undergoing large displacements \cite{Durv03,Durv05}, we present here complete simulations of small
samples of textile composites, in which all fibers involved in the fabric are considered. The main
advantage of this approach is that it offers a direct view to phenomena taking place at the level of
fibers. Since all components are taken into account, there is furthermore no need to identify
behaviours of intermediate models.

The core of our approach is in the detection and modelling of contact-interaction within a large
number of fibers. The principles of this modelling are briefly recalled in the first section.
Adapted algorithms enables efficient solutions of nonlinear problems comprising up to 30 000 contact
elements. The next section is devoted to the computation of the initial configuration of the fabric,
which is actually an unknown of the problem. Starting from a theoretical arrangement of fibers and
yarns, this configuration is computed by progressively enforcing given contact conditions at
crossings. The way the elastic coating is meshed and coupled with the fibers of the fabric by the
means of coupling elements is then presented. Numerical results for biaxial tensions and shear
loadings are finally given for plain weave and twill samples. They demonstrate the ability of the
model to reproduce the nonlinear behaviour of textile structures.

\section{Taking into account of contact-friction between fibers}

\subsection{Modelling aspects}

The automatic detection and taking into account of contact-friction is the core of the model for the
textile composites. Let us recall briefly some principles. Many difficulties are encountered in the
task of detecting contact between fibers : we assume the fibers can undergo large relative
displacements, that contact may appear or disappear anywhere, and at any time.  Our detection of
contact between fibers is based on the construction, in each region where two parts of fibers are
close enough, of an intermediate geometry, defined as an average of the two geometries of the close
parts of fibers. This intermediate geometry is assumed to approximate the actual possible contact
geometry. It is used to provide with a contact search direction which plays a symmetrical part with
respect to the two opposite structures, and as a geometrical support for an independant
discretization of the contact problem. Contact elements are created, on this intermediate geometry,
by associating two material particles, belonging to both fibers in interaction, that are predicted
to enter into contact at a given point on the intermediate geometry. Particles making up these
contact elements may be located anywhere on the beam elements of the fibers, and not only at nodes
or other special integration points. This offers a great flexibility to the method and allows to
position contact elements very accurately, for instance at the crossing between fibers. This process
of determination of contact elements is a predictive and non linear process, since it depends on the
solution itself. For these reasons, contact elements are regularly regenerated during the solution
process.

\subsection{Algorithmical and numerical aspects}

In structures involving a large number of fibers, one has also to deal to numerous contact elements,
and numerical models need therefore to be very efficient to solve the global nonlinear problem in a
reasonable number of iterations. To improve the robustness and efficiency of contact algorithms, we
use a regularized penalty method to control the penetration gap at each contact element. For very
small penetrations, we consider the normal reaction varies quadratically with the gap, which ensures
a zero derivative, and a continuous contact stiffness when the gap goes to zero. This stabilizes
largely the algorithm. The second point which has been very important to improve the global
algorithm, was to adapt locally, that is in each contact zone, the penalty coefficient, in such a
way that the maximum penetration in each contact region is controlled by a given threshold. When a
fixed penalty coefficient is considered, the penetration at contact element depends on the forces
exerted by the interacting structures. But these forces can vary very largely within the global
assembly of fibers, and also with the loading conditions. In this situation, keeping the penalty
coefficient fixed will give penetrations of very different amplitudes. The local adaptation of the
penalty coefficient to keep the maximum penetration of the same order in the different contact
regions seems to be a good guaranty to have the contact algorithm converge in the best way in all
regions. Thanks to these improvements of the contact algorithms, in the numerical examples presented
in the following, cases with about 20.000 contact elements could be solved with only about 60
iterations per loading step.

\section{Computation of the initial configuration of the fabric}

The initial configuration of the fabric is an \textit{a priori} unknown of the problem. All we have
access to is the theoretical arrangement of fibers in individual yarns, and the relative positions
of yarns at crossings in the weaving, depending on the weaving structure.

The computation of this initial configuration is one of the contributions of the simulation code,
and provides with significant informations, concerning especially the shapes of the cross-sections
of the yarns in the fabric. To obtain this initial configuration, we start from a theoretical
configuration, which follows only the relative arrangement of fibers within yarns, and where yarns
penetrate each other at crossings. The equilibrium configuration of the woven structure is then
progressively reached, simply by enforcing, during first steps, the direction of contact between
fibers of different yarns at the crossings between yarns, depending on the weave pattern. In order
to have only small perturbations between steps, the contact between fibers of different yarns is
searched, during this initialization stage, only at small distances. Once the weaving pattern is
checked at each crossing, that is all fibers of one yarn are above, or below, the fibers of the
other yarn, the enforcement of the contact directien is left, and the classical contact conditions
are considered between all fibers. Results at different steps of this initialization stage may be
seen on Figure \ref{fig:FabricConfIni}.

The goal of this initialization stage is to get the right geometry for each fiber of the fabric. The
configuration obtained at the end of this stage is then taken as the initial configuration for the
rest of the computation, in such a way that stresses (tractions) are considered to be zero in the
following.

\section{Consideration of an elastic matrix : coupling with fibers of the fabric}

The consideration of an elastic matrix is necessary to identify the behaviour of textile
composites. An accurate modelling of the interface between fibers and matrix seems very difficult to
reach because of the complexity of the geometry of the yarns, and of the fineness of discretization
it would induce in the matrix in order to have conforming meshes between the fibers and the
matrix. Assuming components of the fabric are stiffer than the coating material, our choice is to
model the matrix and its coupling with the fabric rather roughly.

\begin{figure}[!h]
   \includegraphics[width=6.2cm]{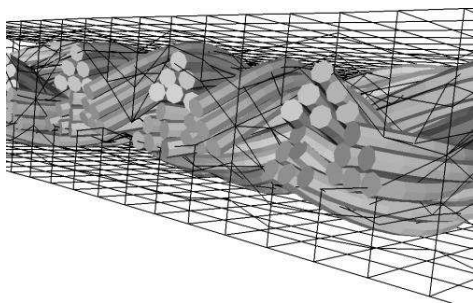}
  \caption{Nonconforming meshes for the elastic matrix and the fibers}
  \label{fig:MatrixMeshDetails}
\end{figure}

The meshing of the matrix and the coupling with fibers of the matrix are carried out in the
following way. Hexahedral elements are created for the matrix, according to a structured pattern, so
that they slightly penetrate the volume of the yarns (see Figure \ref{fig:MatrixMeshDetails}). This
penetration creates an overlapping region where external fibers of the yarns are inside the volume
of finite elements of the matrix. To ensure the mechanical coupling between the fibers and the
matrix, for each beam node of a fiber which is inside an element of the matrix, we create a
``coupling element'' between this node and the particle of the matrix occupying the same position.
This coupling element behaves as an elastic link between the node and the particle, whose stiffness
is equal to the stiffness of the matrix. Due to this adapted stiffness, these coupling elements are
able to couple kinematical fields of different nature, and approximated by very different
discretized fields in the two linked structures.

\section{Numerical results}

\subsection{Computation of initial configuration for plain weave and twill samples}

\begin{figure}[!h]
  \centerline{\includegraphics[width=6.2cm]{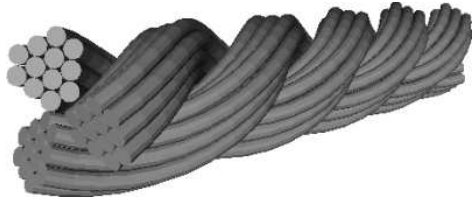}}
  \caption{Start configuration for a 36 fibers yarn}
  \label{fig:YarnConfIni}
\end{figure}

\begin{figure}[!h]
  \centerline{\includegraphics[width=5.7cm]{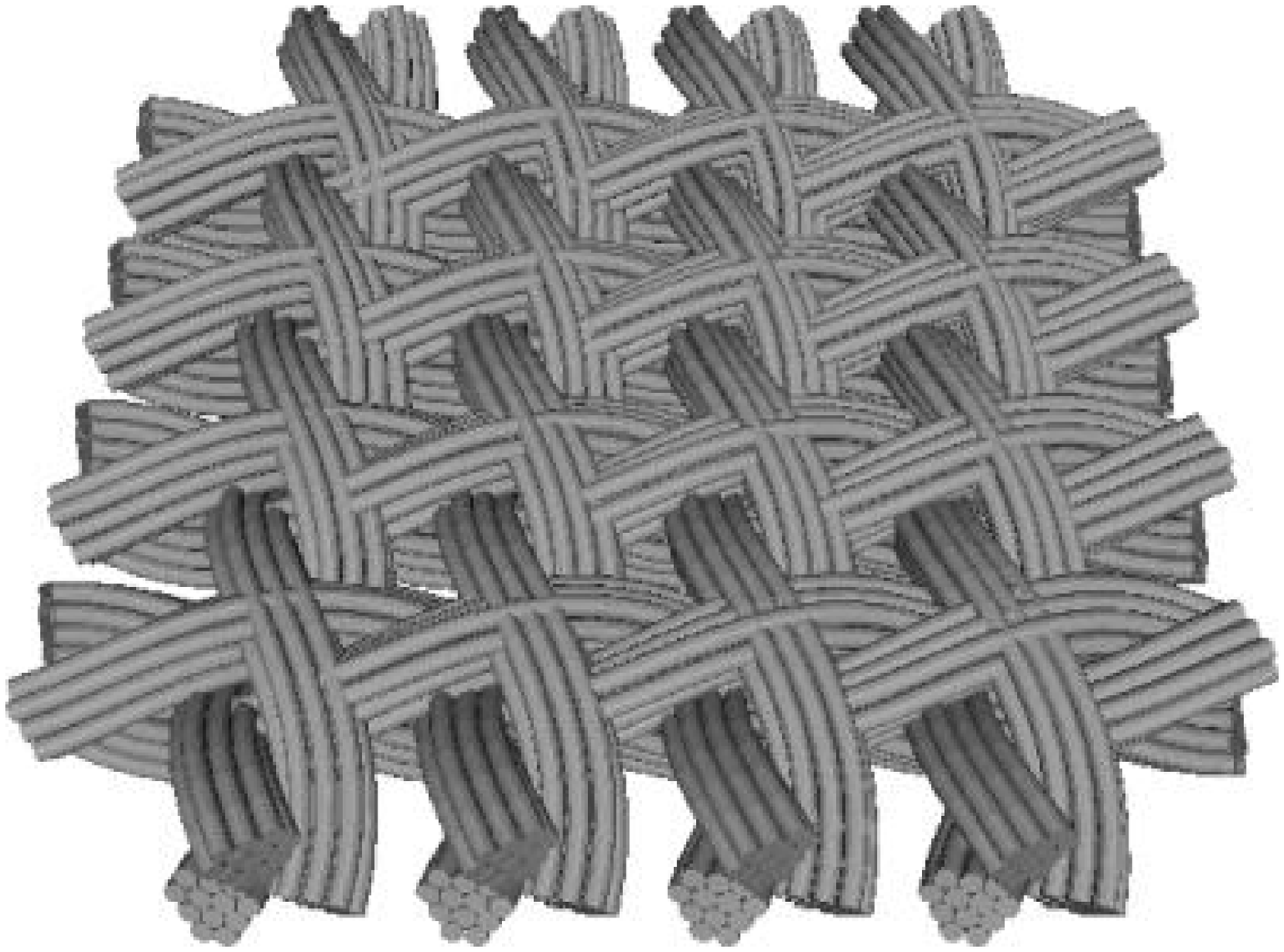} \hspace{1cm} \includegraphics[width=5.7cm]{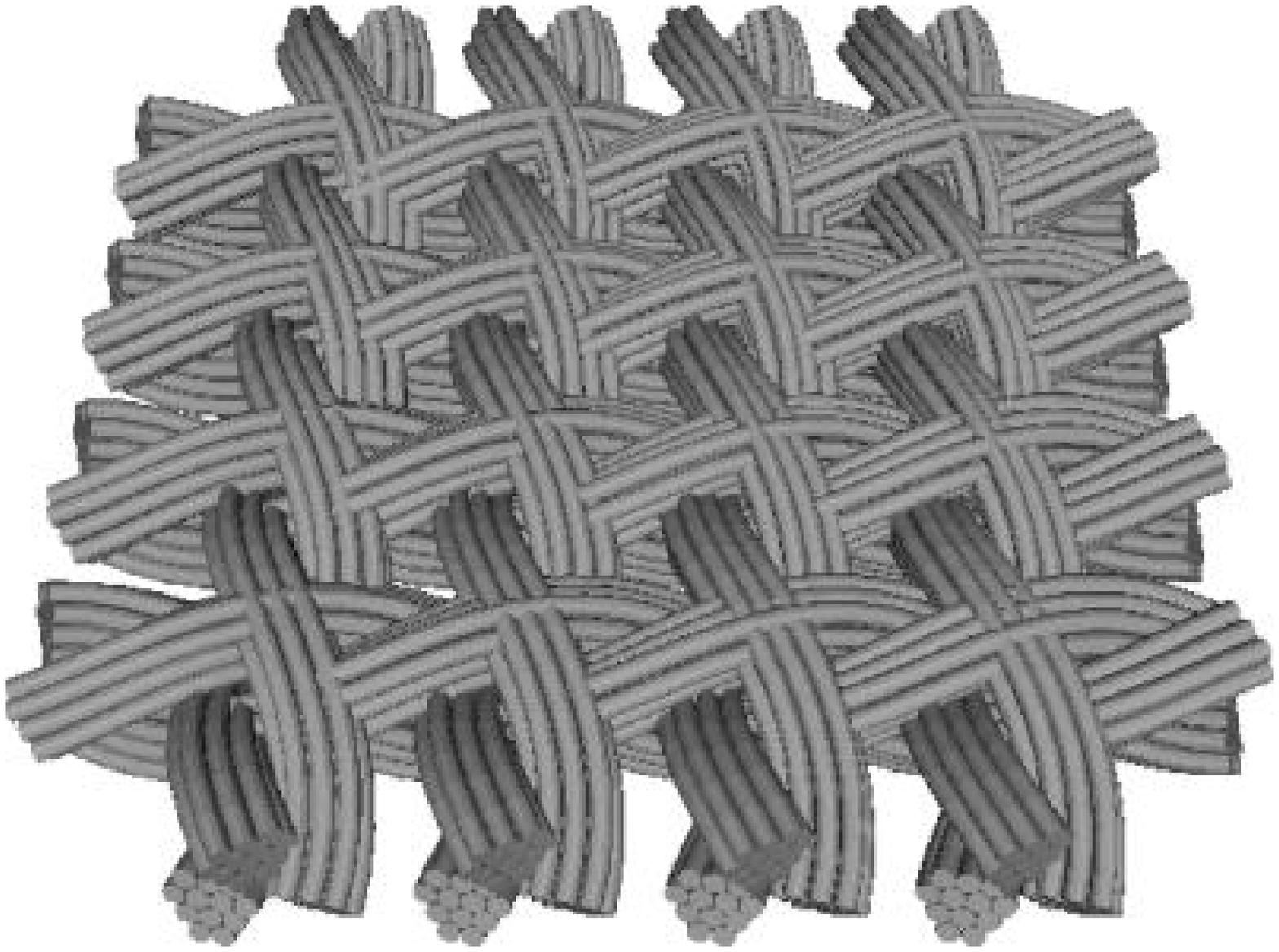}}
  \centerline{\footnotesize{1. Theoretical start configurations}}
  \centerline{\includegraphics[width=5.7cm]{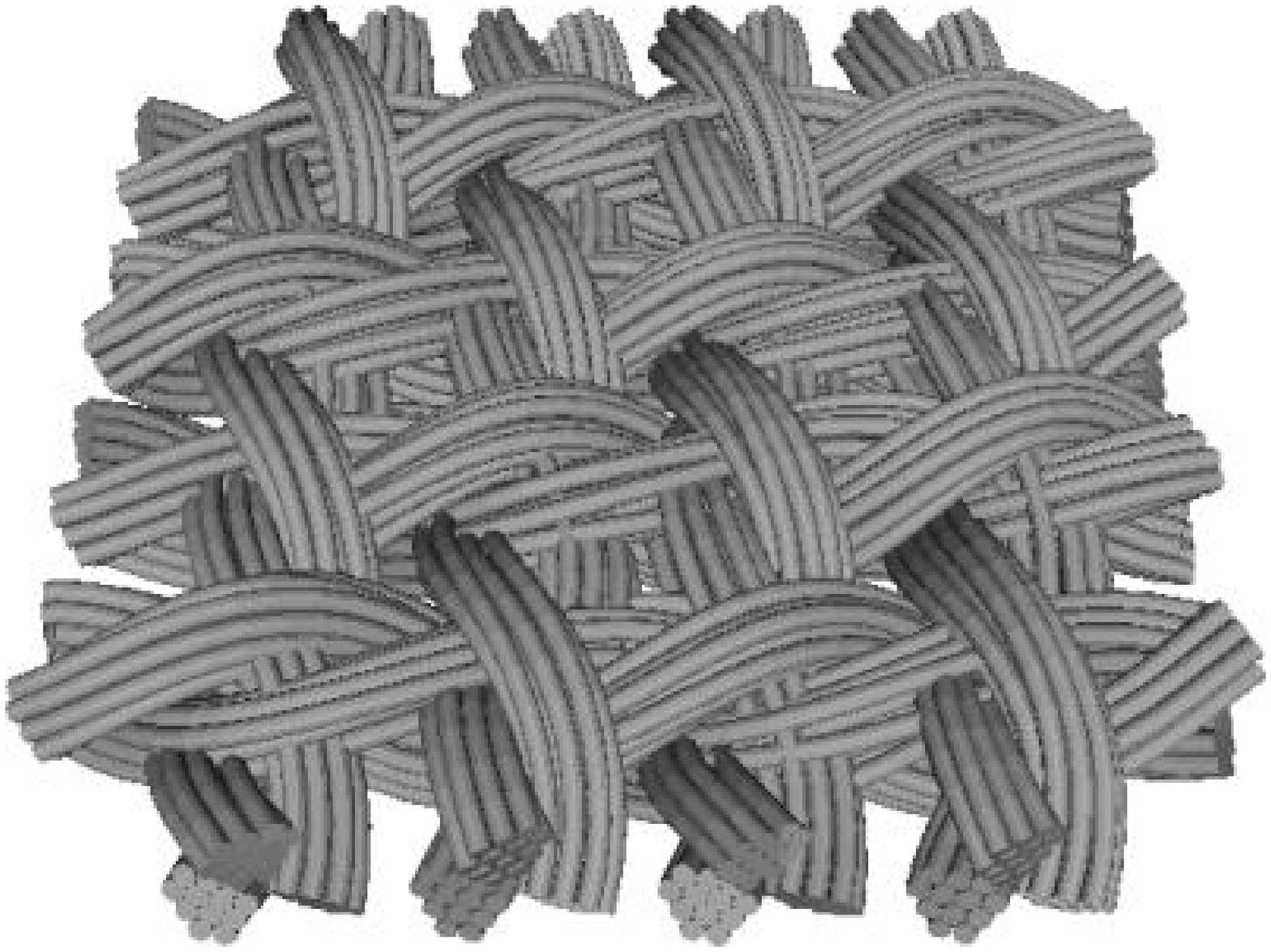} \hspace{1cm} \includegraphics[width=5.7cm]{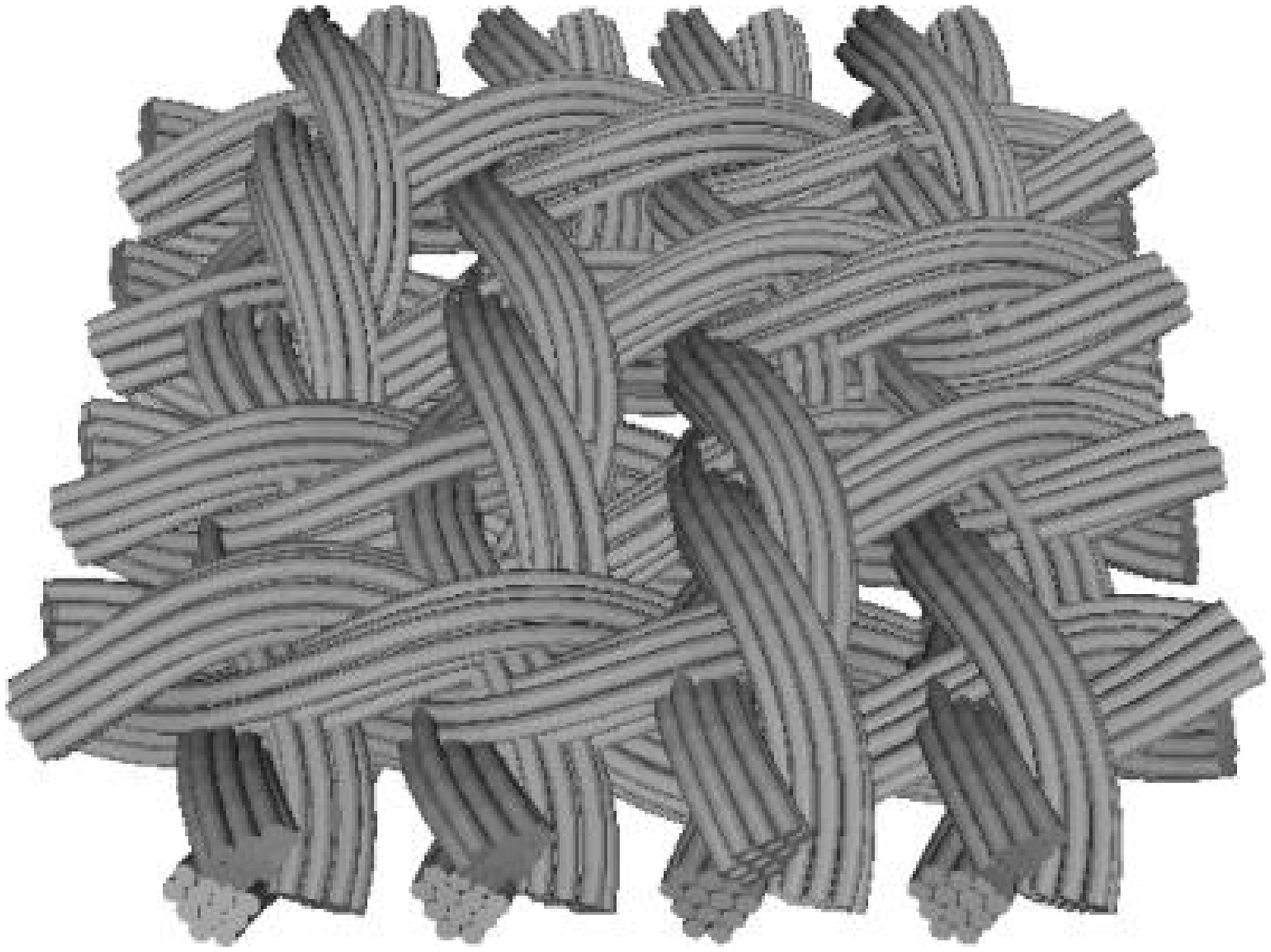}}
  \centerline{\footnotesize{2. Intermediate (unbalanced) configurations}}
  \centerline{\includegraphics[width=5.7cm]{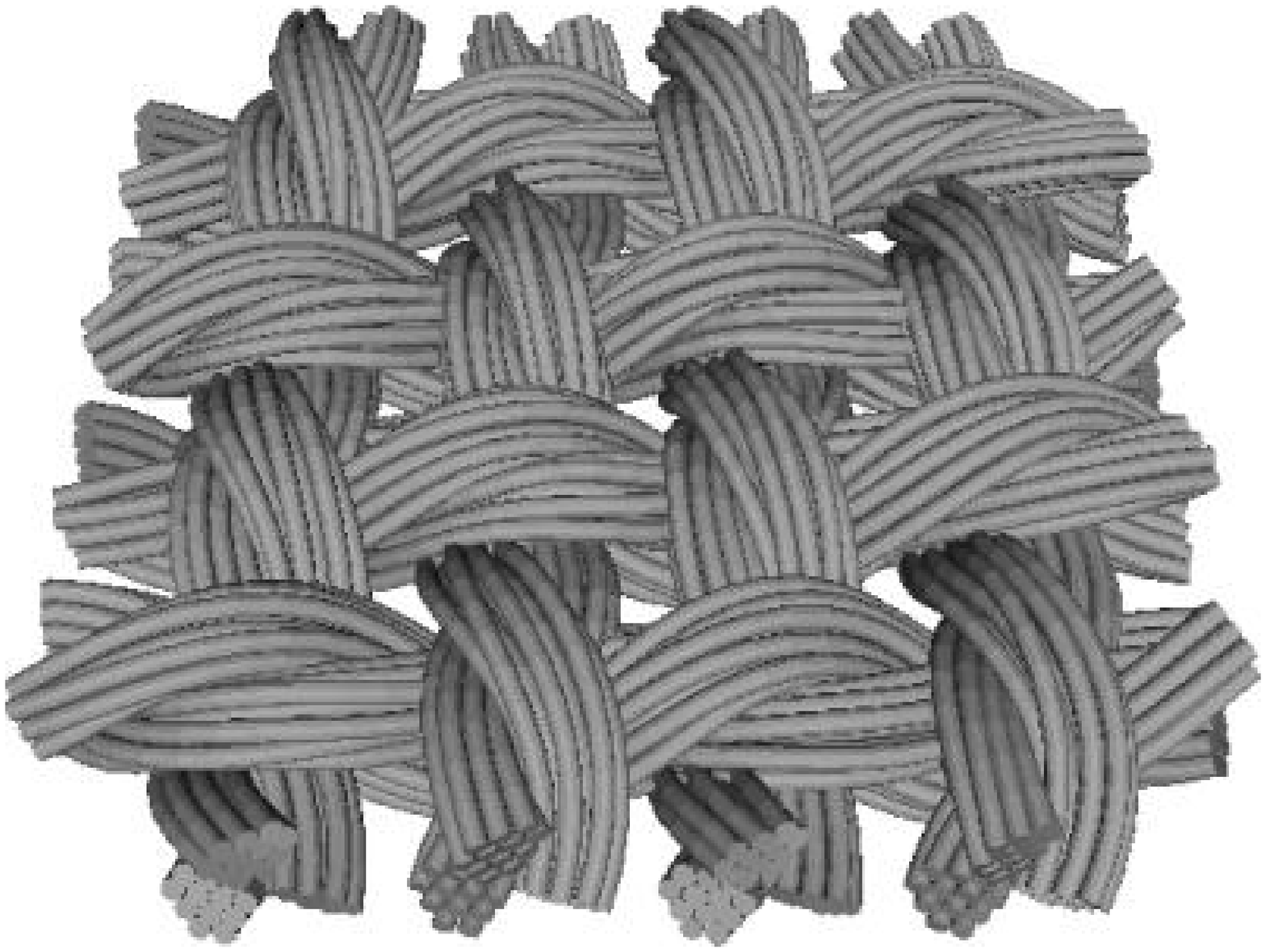} \hspace{1cm} \includegraphics[width=5.7cm]{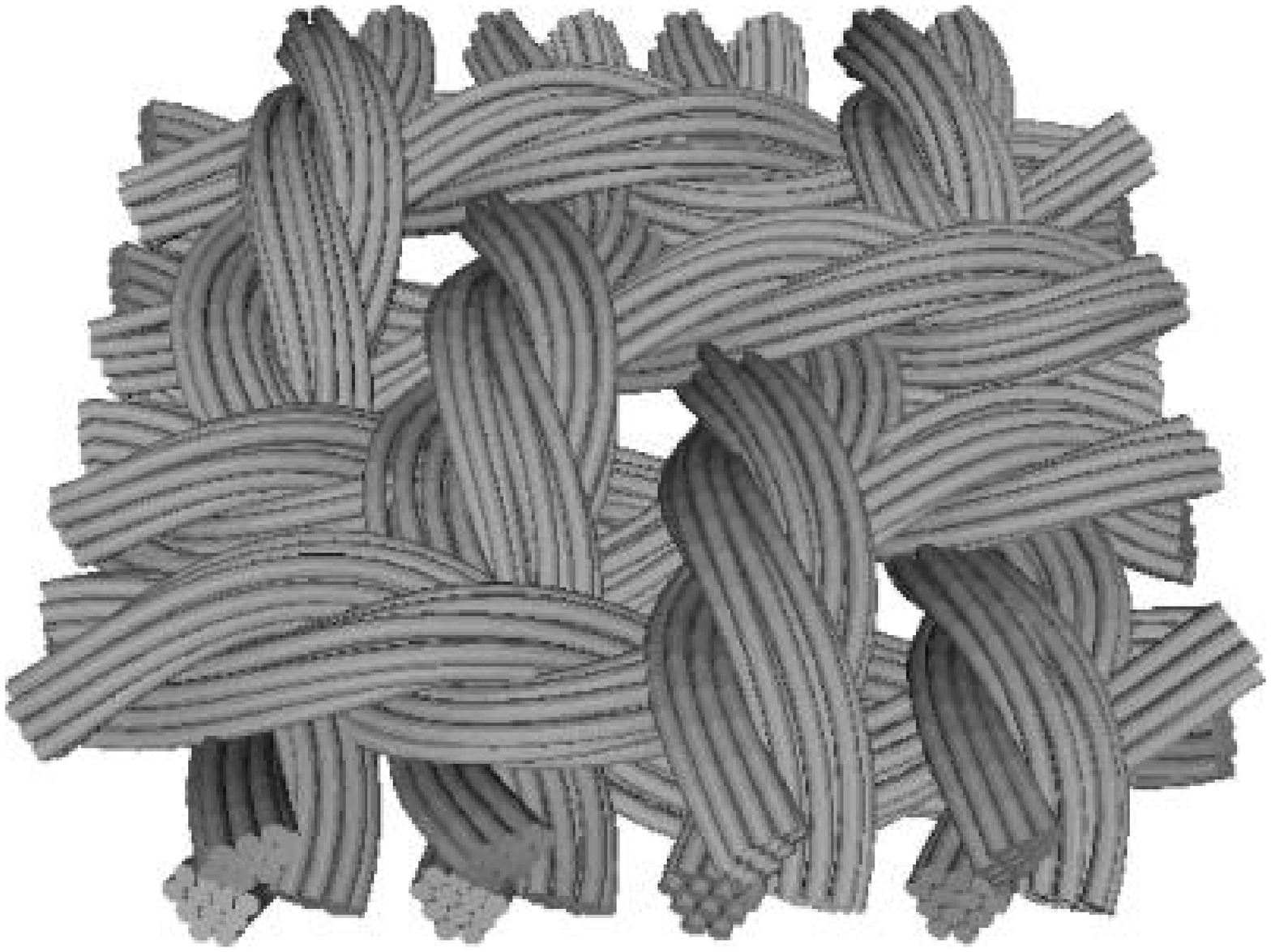}}
  \centerline{\footnotesize{3. Intermediate (unbalanced) configurations}}
  \centerline{\includegraphics[width=5.7cm]{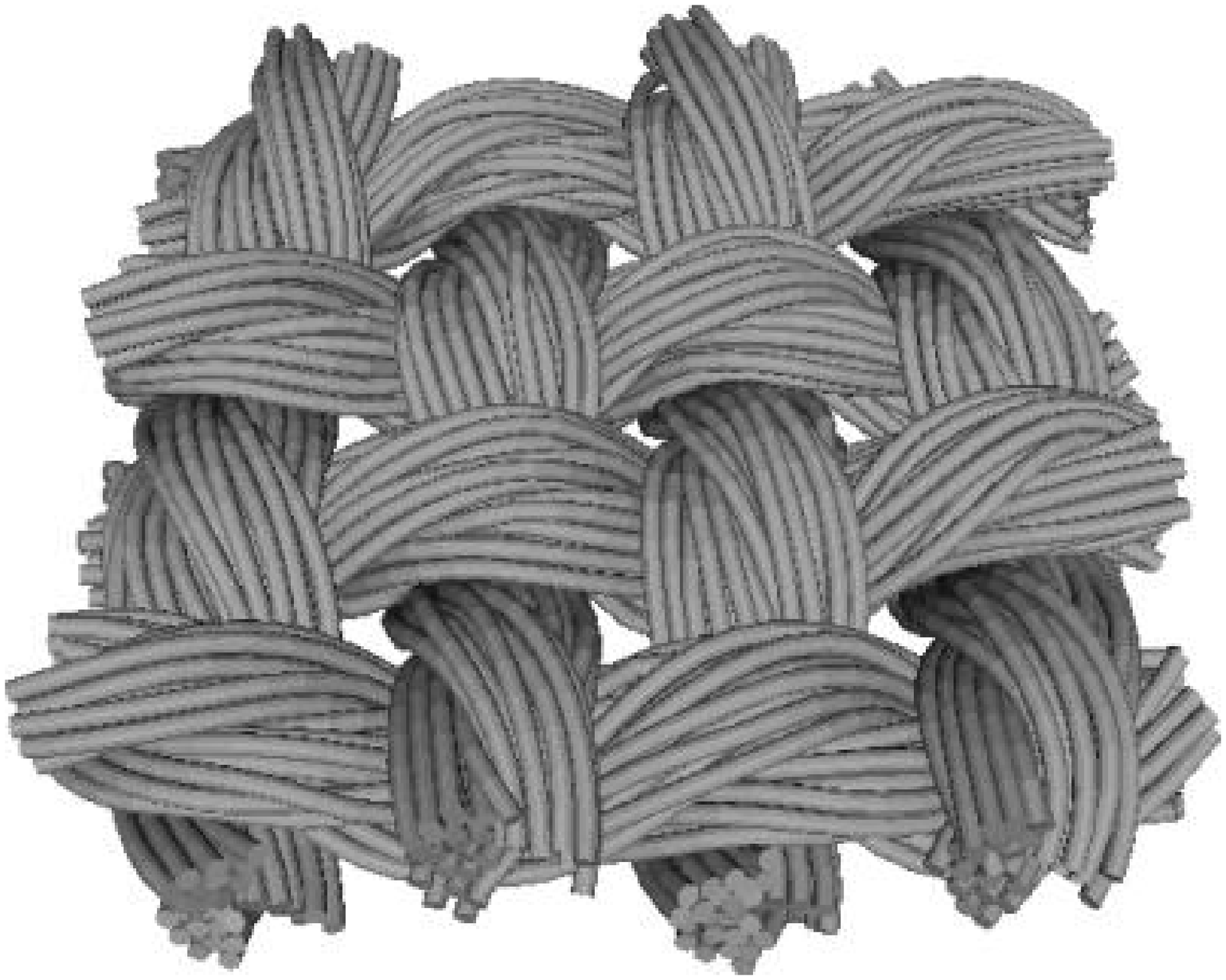} \hspace{1cm} \includegraphics[width=5.7cm]{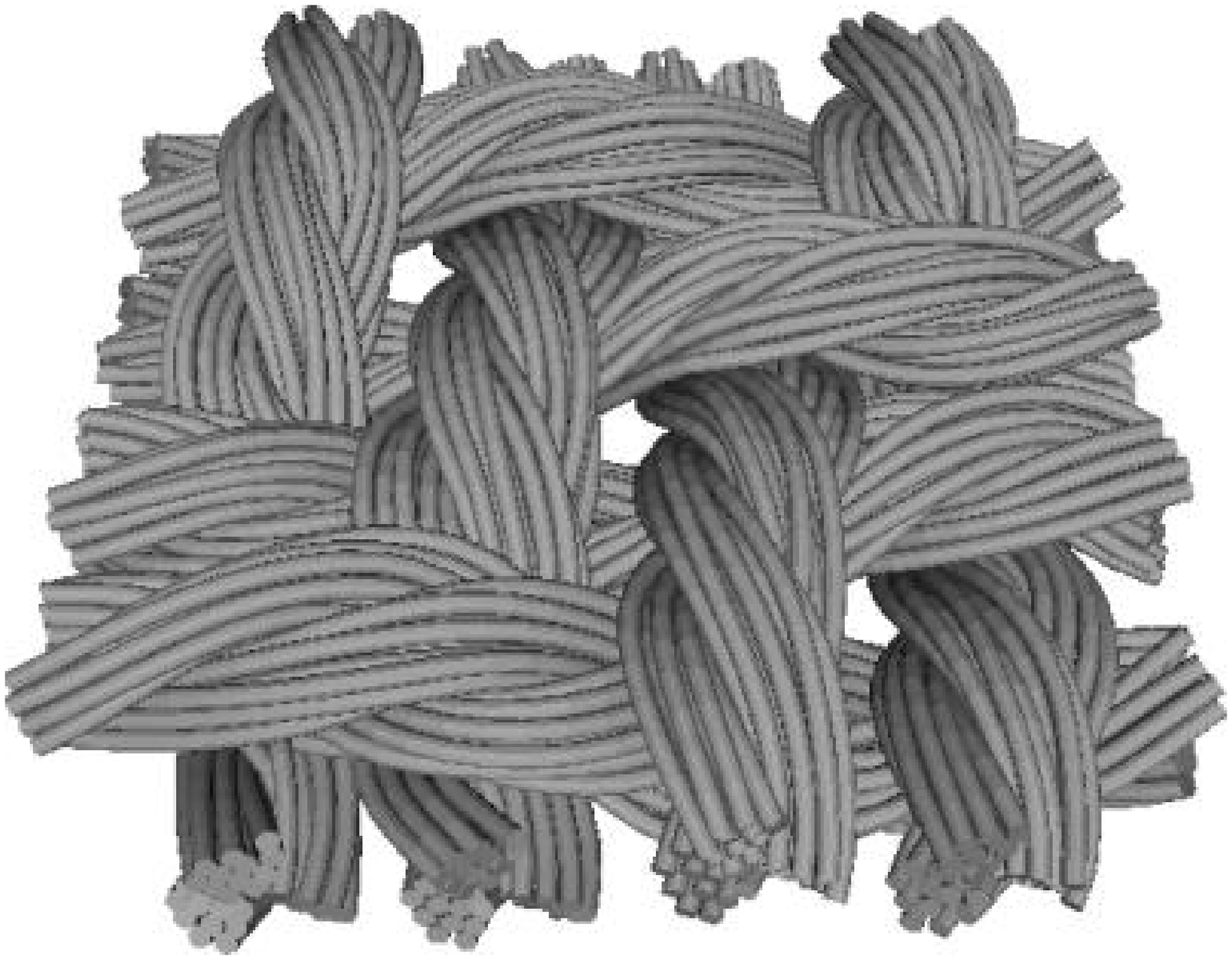}}
  \centerline{\footnotesize{4. Computed balanced configurations}}
  \caption{Progressive calculation of the initial configuration for a plain weave (left) and a twill
    (right) sample}
  \label{fig:FabricConfIni}
\end{figure}

To validate the computation of the initial configuration, we have taken yarns made with 3 bundles of
12 fibers, that is made of 36 fibers each, as illustrated on Fig. \ref{fig:YarnConfIni}.
To keep a reasonable CPU time, we have only considered samples made of 8 interlocked yarns, and
arranged according to two different weave structures : a plain weave, and a 2x2 twill. The two
models comprise around 95 000 dofs. In the final step, about 25 000 contact elements where built and
taken into account in the computation, among which, about 18 000 were active. Successive
configurations obtained for the two weavings are shown on Figure \ref{fig:FabricConfIni}.

\begin{figure}[!h]
  \centerline{\includegraphics[width=6.5cm]{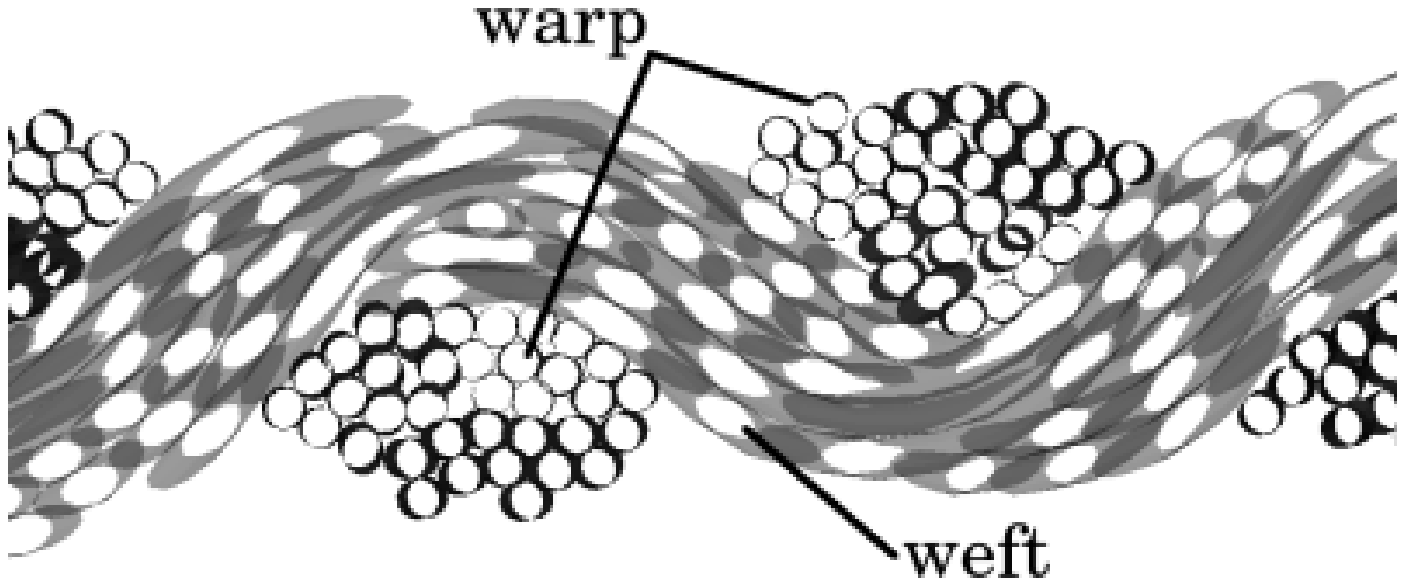} \hspace{1cm} \includegraphics[width=6.5cm]{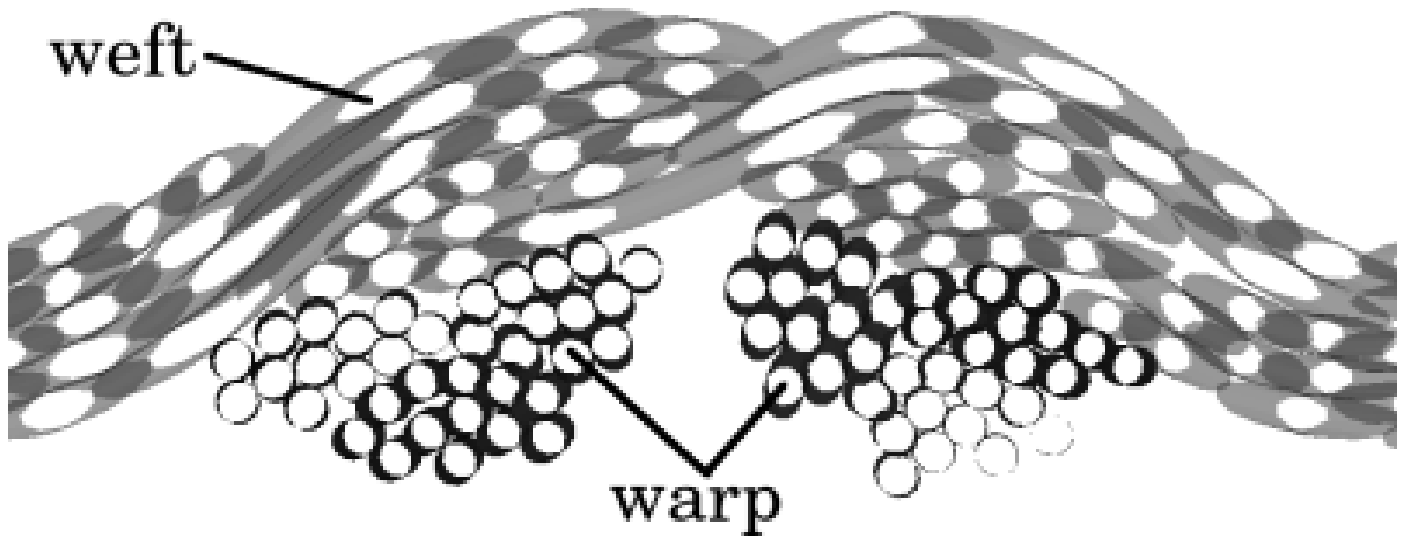}}
  \caption{Cutouts of the deformed configurations at the last step of the initialization for a plain
  weave (left) and a twill (right)}
  \label{fig:CutOutConfIni}
\end{figure}

\clearpage

\begin{figure}[!h]
  \centerline{\includegraphics[width=5.1cm]{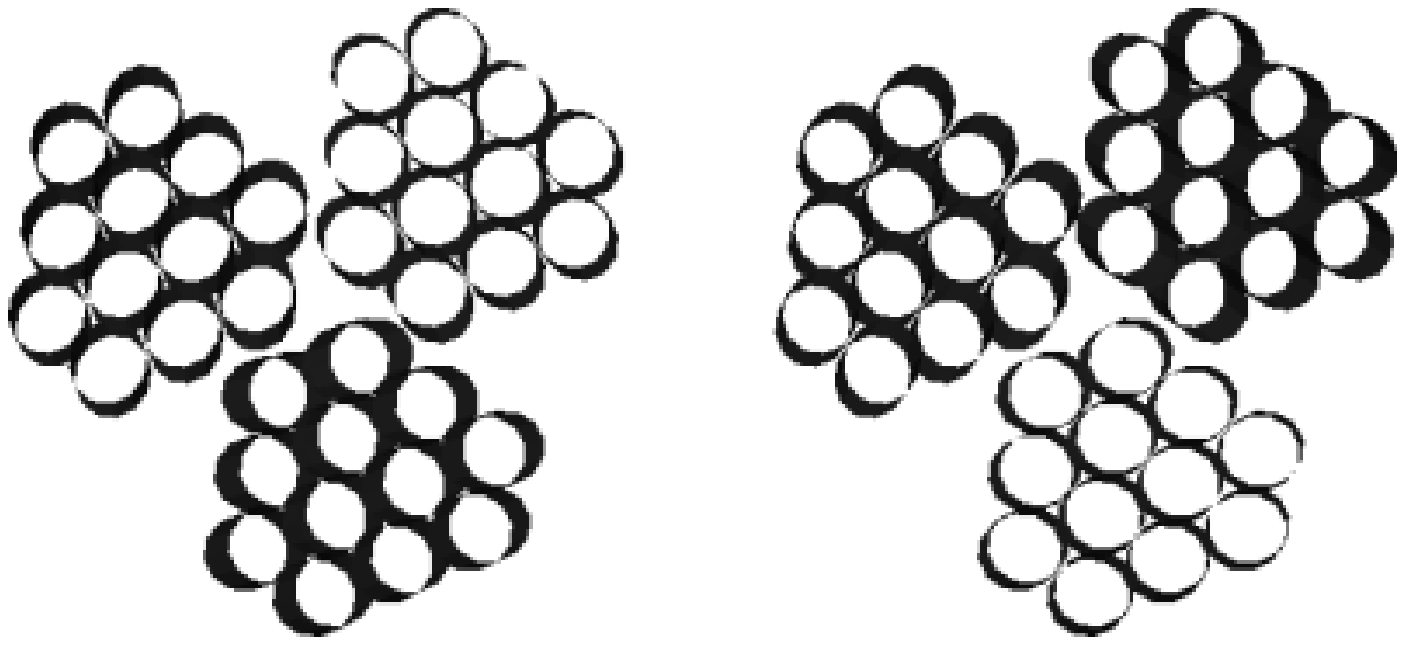} \hspace{2cm} \includegraphics[width=5.1cm]{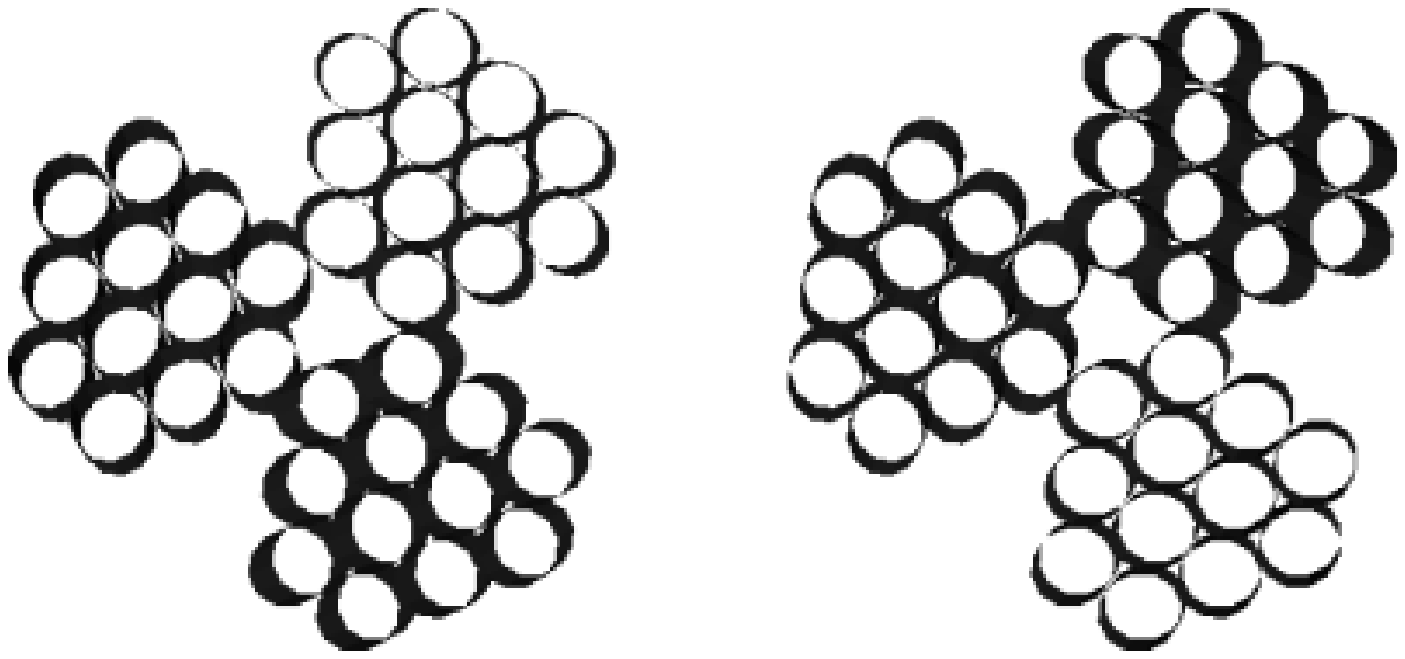}}
  \centerline{\footnotesize{1. Theoretical start configurations}}
  \centerline{\includegraphics[width=6.5cm]{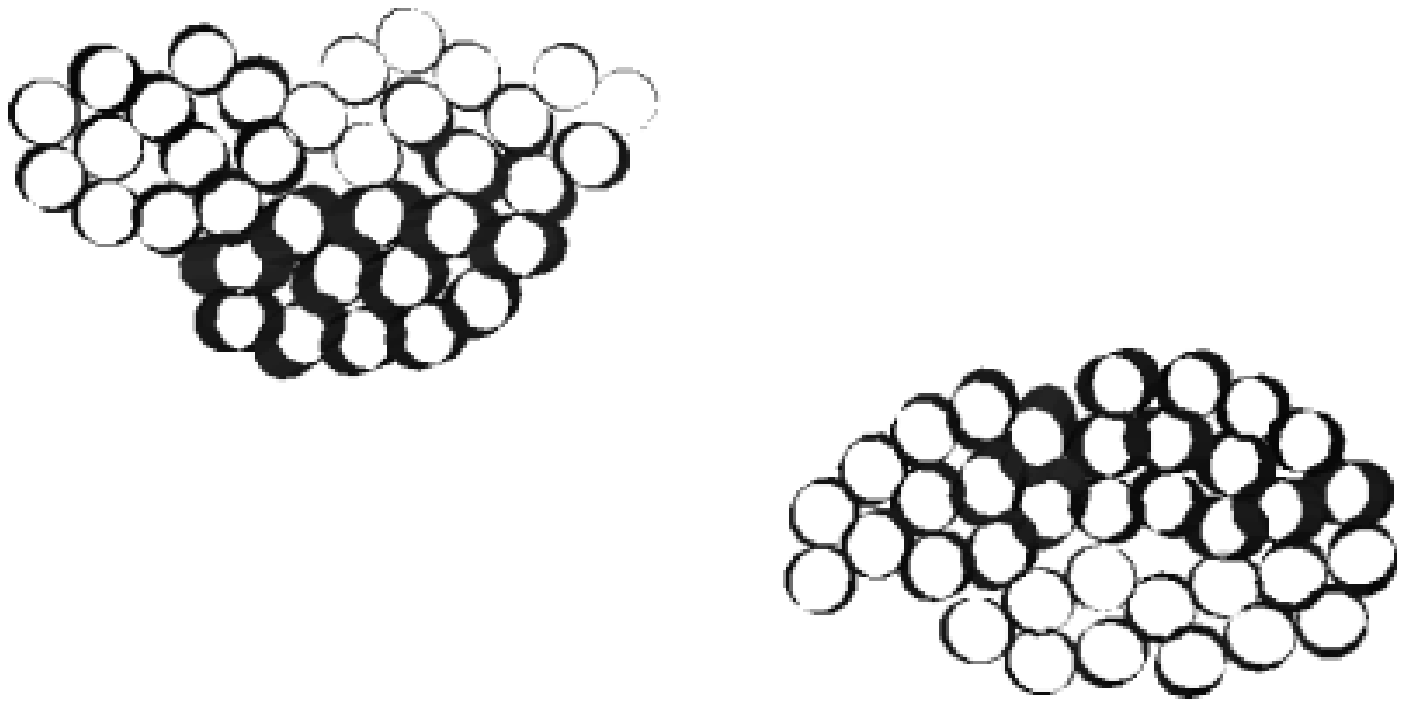} \hspace{2cm} \includegraphics[width=6.5cm]{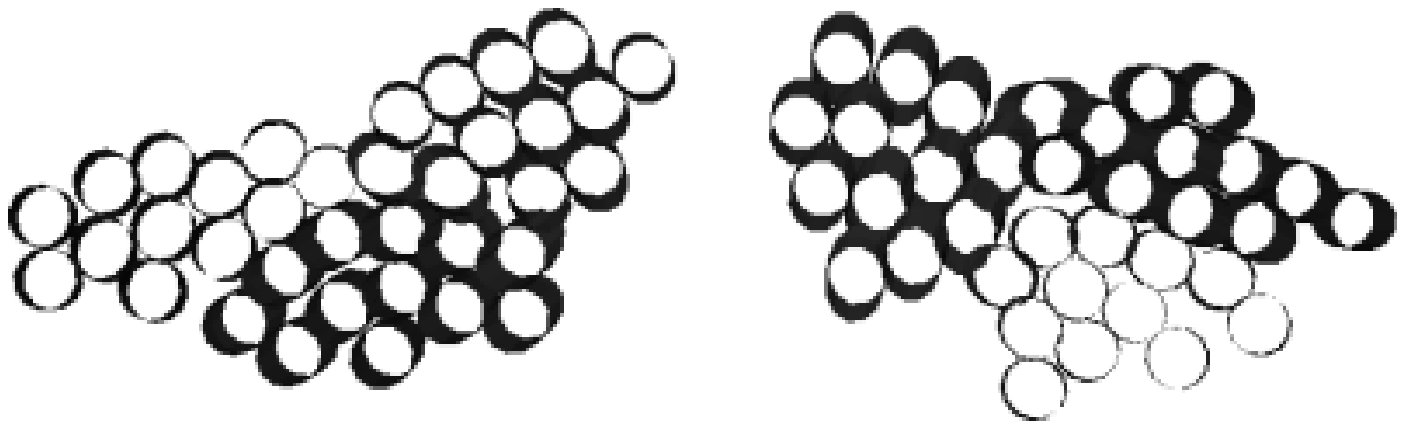}}
  \centerline{\footnotesize{2. Computed balanced configurations}}
  \caption{Evolution of two neighbouring warp yarns cross sections between the theororetical start configuration and
    the computed initial configuration for the plain weave (left) and the twill (right) samples}
  \label{fig:YarnCrossSectConfIni}
\end{figure}

Valuable informations about the geometries of cross-sections of yarns are obtained from this
computation of the initial configuration. Figure \ref{fig:CutOutConfIni} shows the very different
shapes for both the medium line of yarns and their cross-sections in function of the two weave
structures. Figure \ref{fig:YarnCrossSectConfIni} enlightens the great variations in the
arrangements of fibers in yarns between the start configuration and the computed initial
configuration of the two fabrics.

\subsection{Identification of the constitutive behaviour under biaxial tensions and shear loadings}

\subsubsection{Presentation of the models}

\begin{figure}[!h]
  \centerline{\includegraphics[width=8cm]{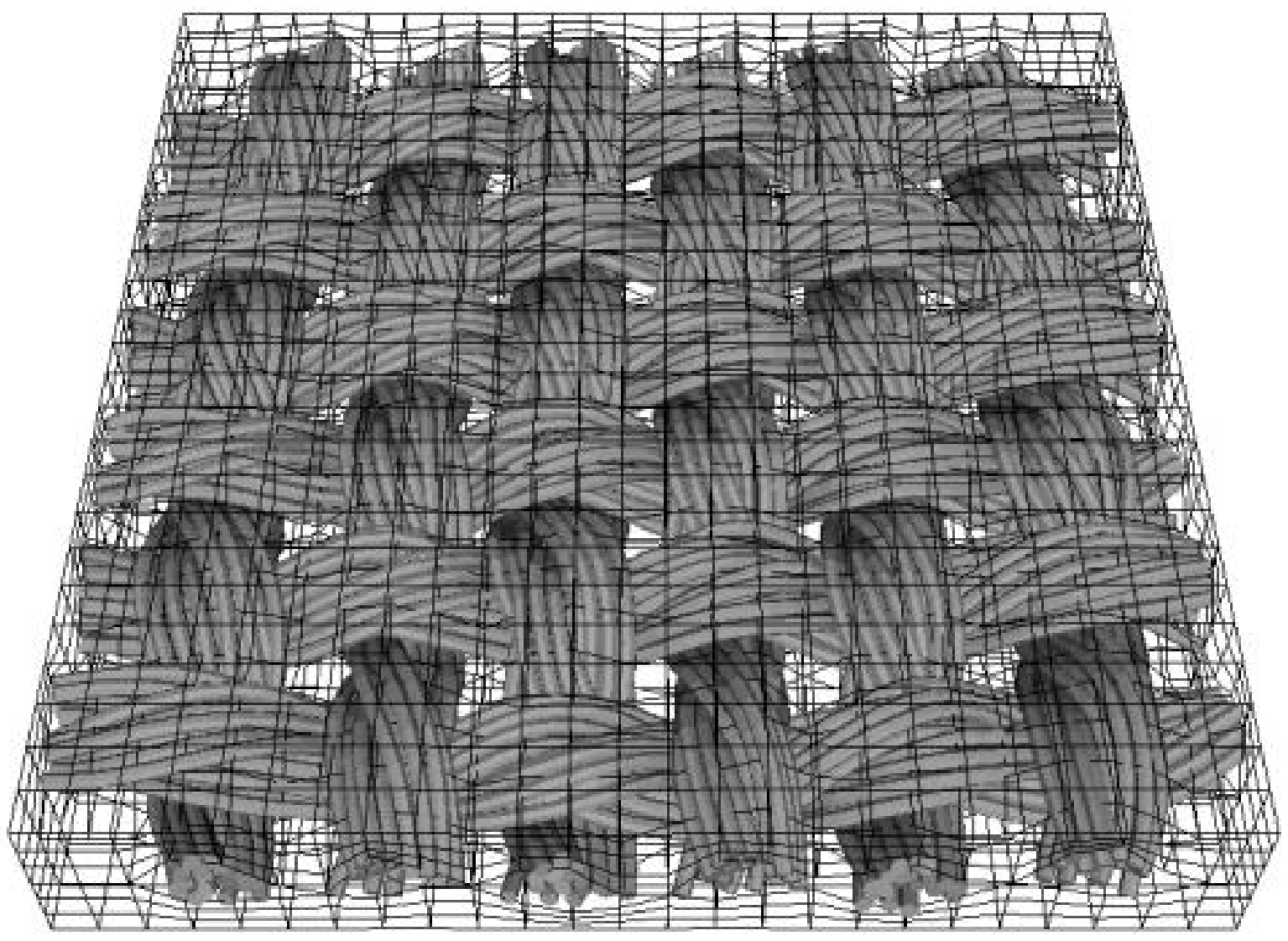} \includegraphics[width=8cm]{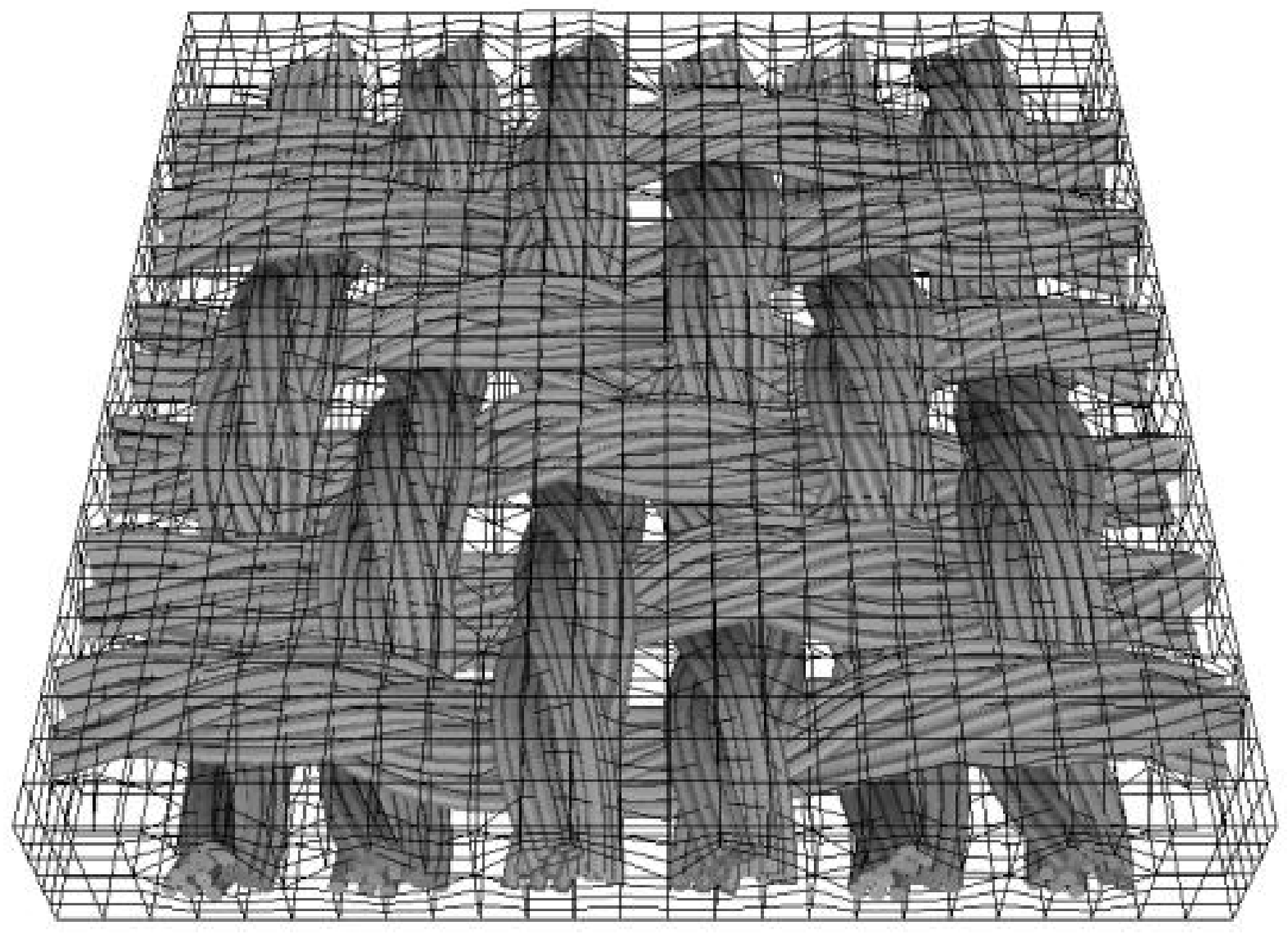}}
  \caption{Initial meshes for the two samples : plain weave (left) and twill (right)}
  \label{fig:ComposMeshes}
\end{figure}

After the calculation of the initial configuration, a restart is done, considering now the presence
of an elastic coating and its coupling with the fibers of the fabric. For this identification, the
studied samples were made of 12 interlocked yarns, each of them comprising 3 bundles of 6 fibers.
Meshes for the global composite structure for plain weave and twill are shown on Figure
\ref{fig:ComposMeshes}.

The ratio between the stiffness of the matrix and the stiffness of the fibers is taken to 1000, and
a coefficient of 0.1 is considered between fibers.

The models have about 120 000 degrees of freedom and about 20 000 contact elements, among which
about 15 000 are active.

\subsubsection{Biaxial tests}

\begin{figure}[!h]
  \centerline{\includegraphics[width=8cm]{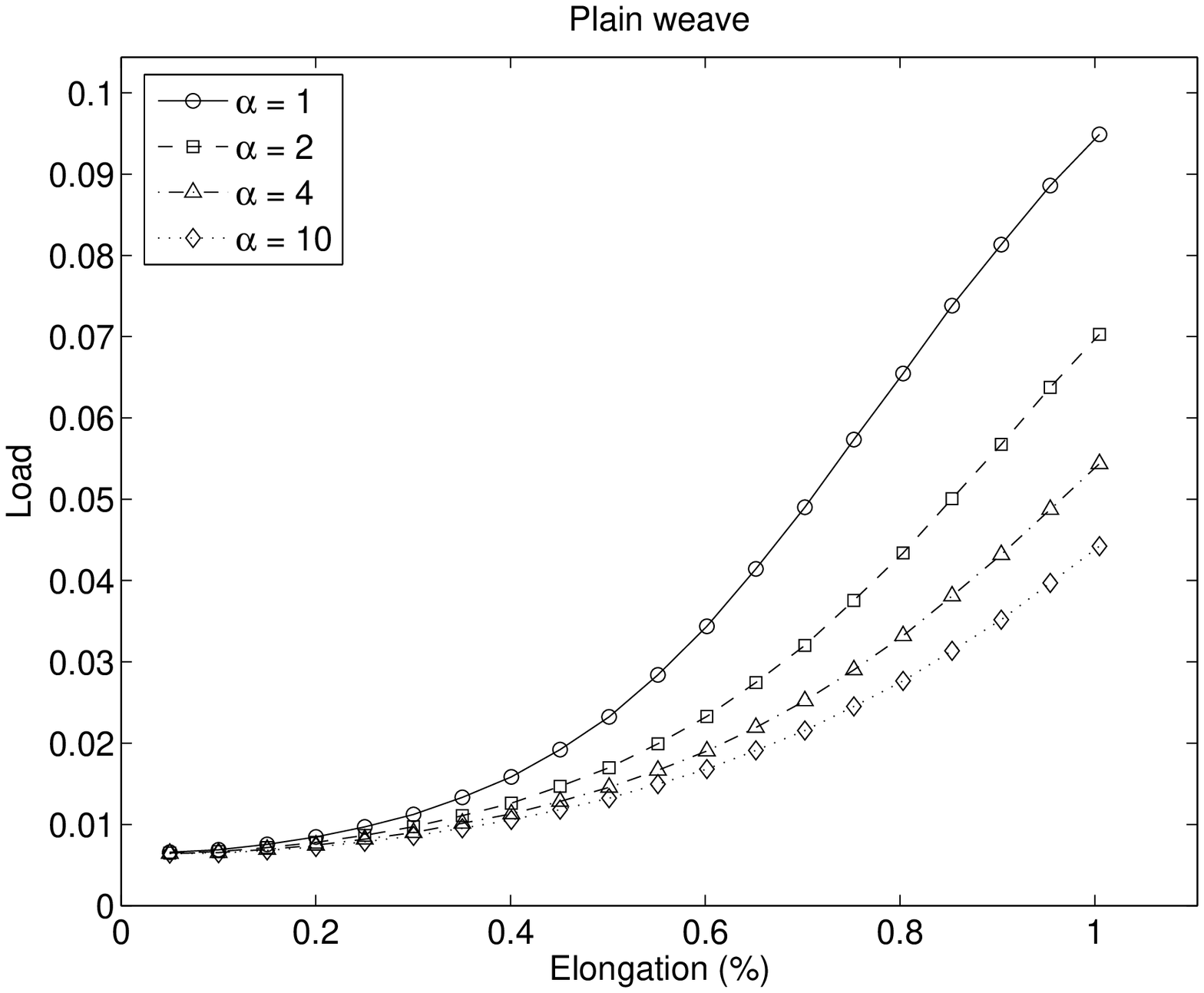} \includegraphics[width=8cm]{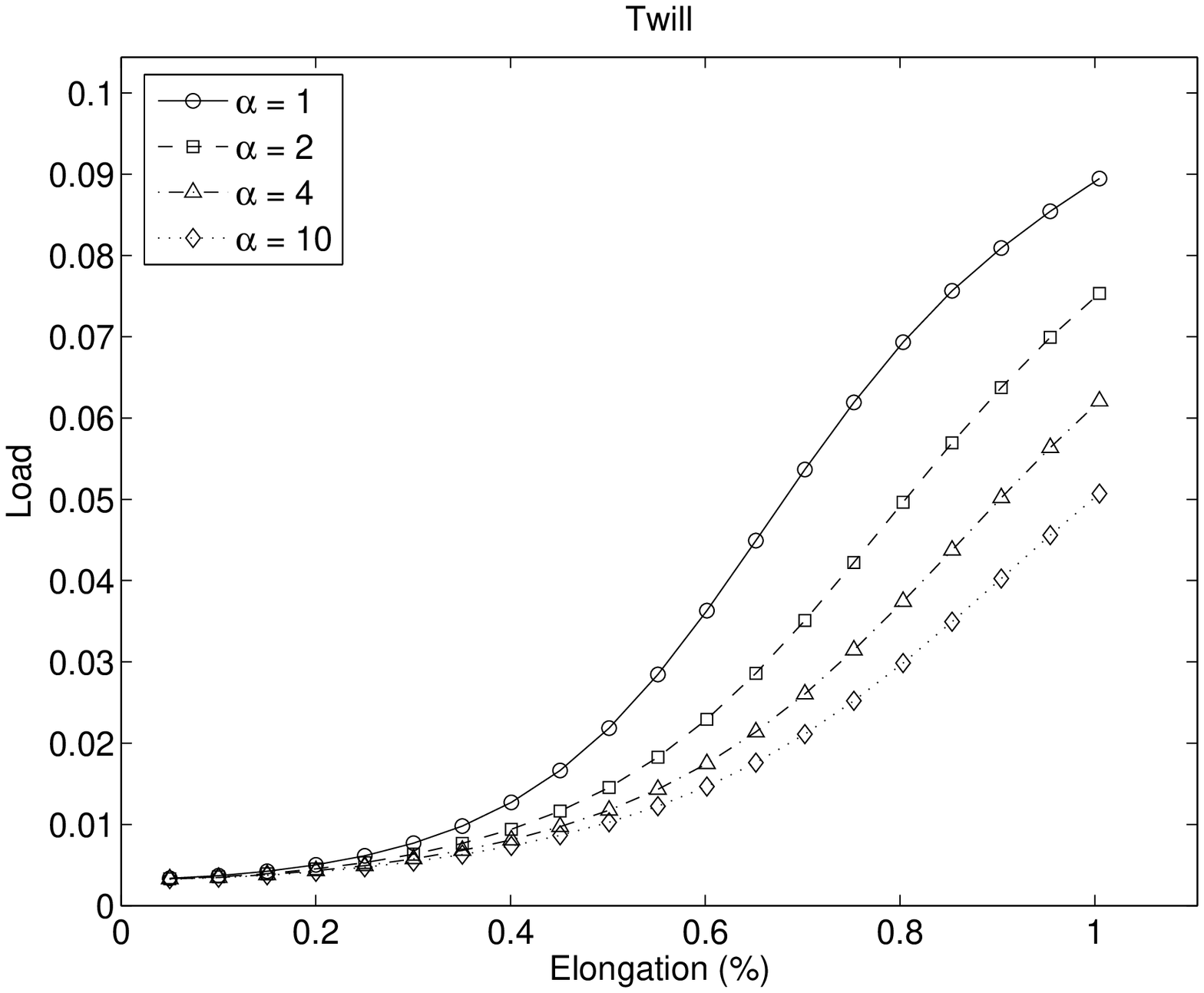}}
  \caption{Biaxial tension loading curves obtained for different ratio between the strains applied
    to warp and weft directions}
  \label{fig:BiaxCurves}
\end{figure}

Biaxial tension simulations have been carried out on the two samples. Different strains have been
prescribed in the warp and weft directions ; the ratio between the steains in these two directions,
denoted $\alpha$, has been taken equal to 1, 2, 4 and 10. The loading curves (Figure
\ref{fig:BiaxCurves}) show well-known nonlinear characteristics at the start of the loading.

\subsubsection{Shear tests}

\begin{figure}[!h]
  \centerline{\includegraphics[width=8cm]{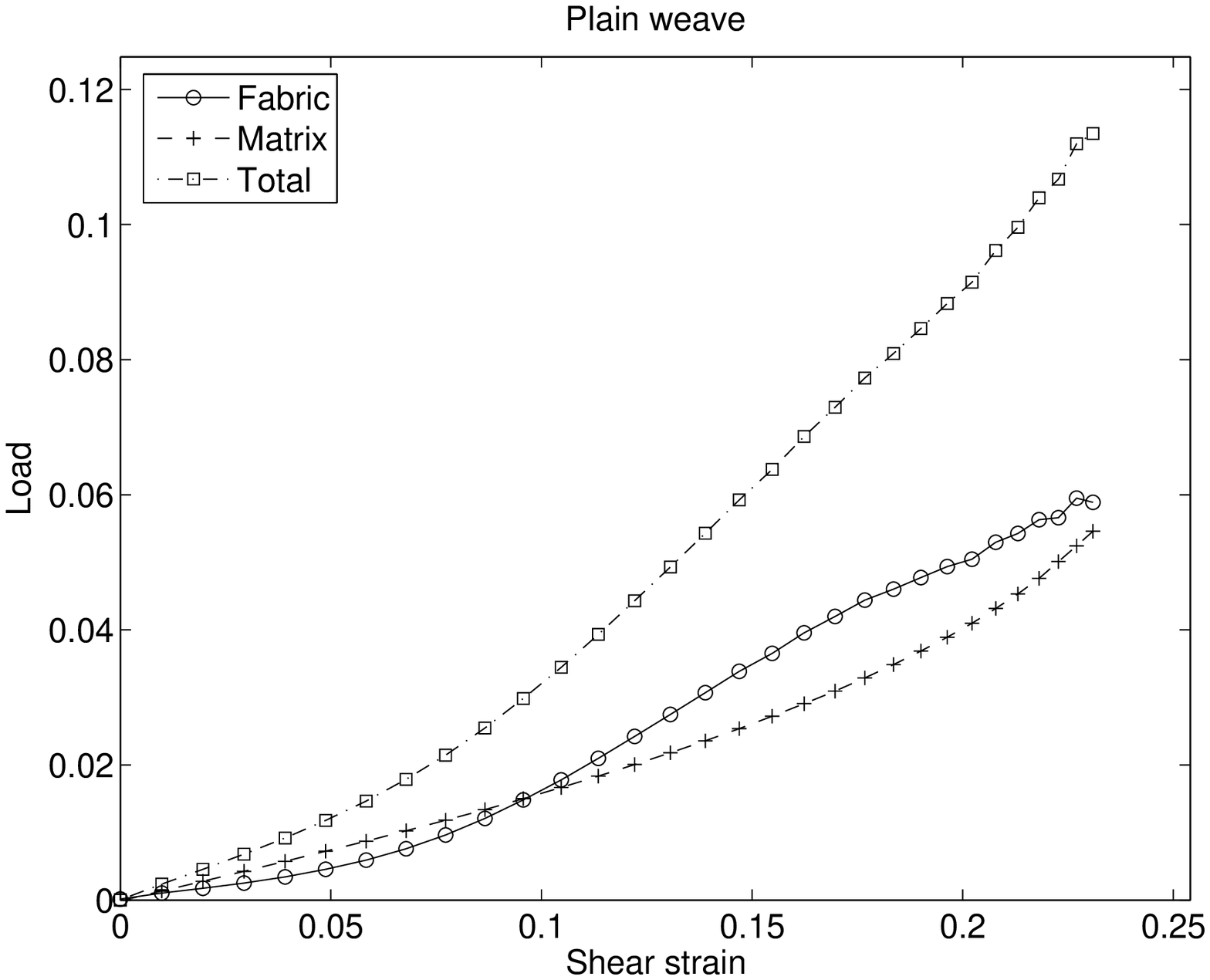} \includegraphics[width=8cm]{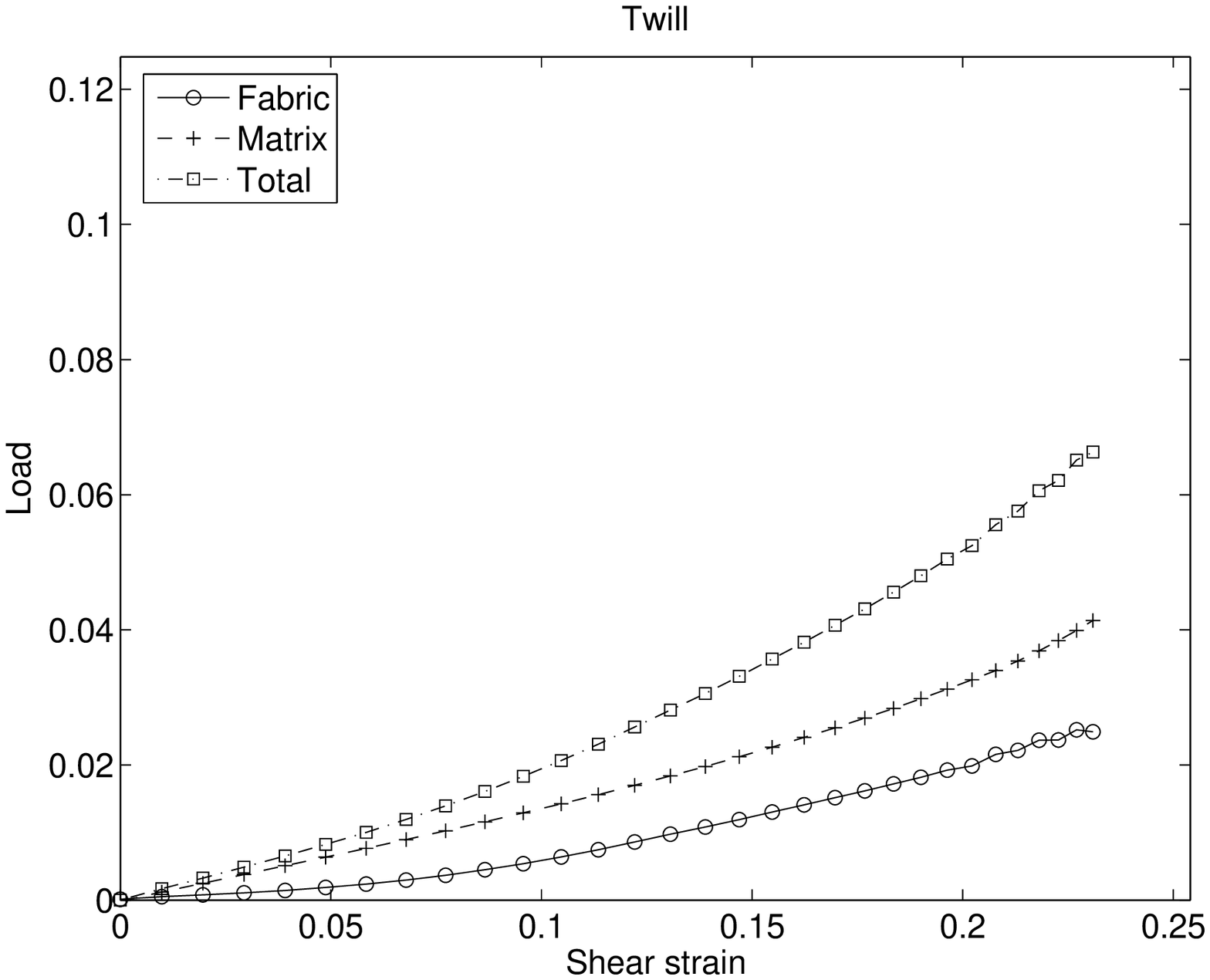}}
  \caption{Shear loading curves}
  \label{fig:ShearCurves}
\end{figure}

Shear loading simulation have been carried out by applying a rotation on the sides of the samples.
In this case, the contribution of the matrix to the total force is not neglectible as shown on the
loading curves (Figure \ref{fig:ShearCurves}). Nonlinar effects are more difficult to analize.

\begin{figure}[!btp]
  \centerline{\includegraphics[width=8cm]{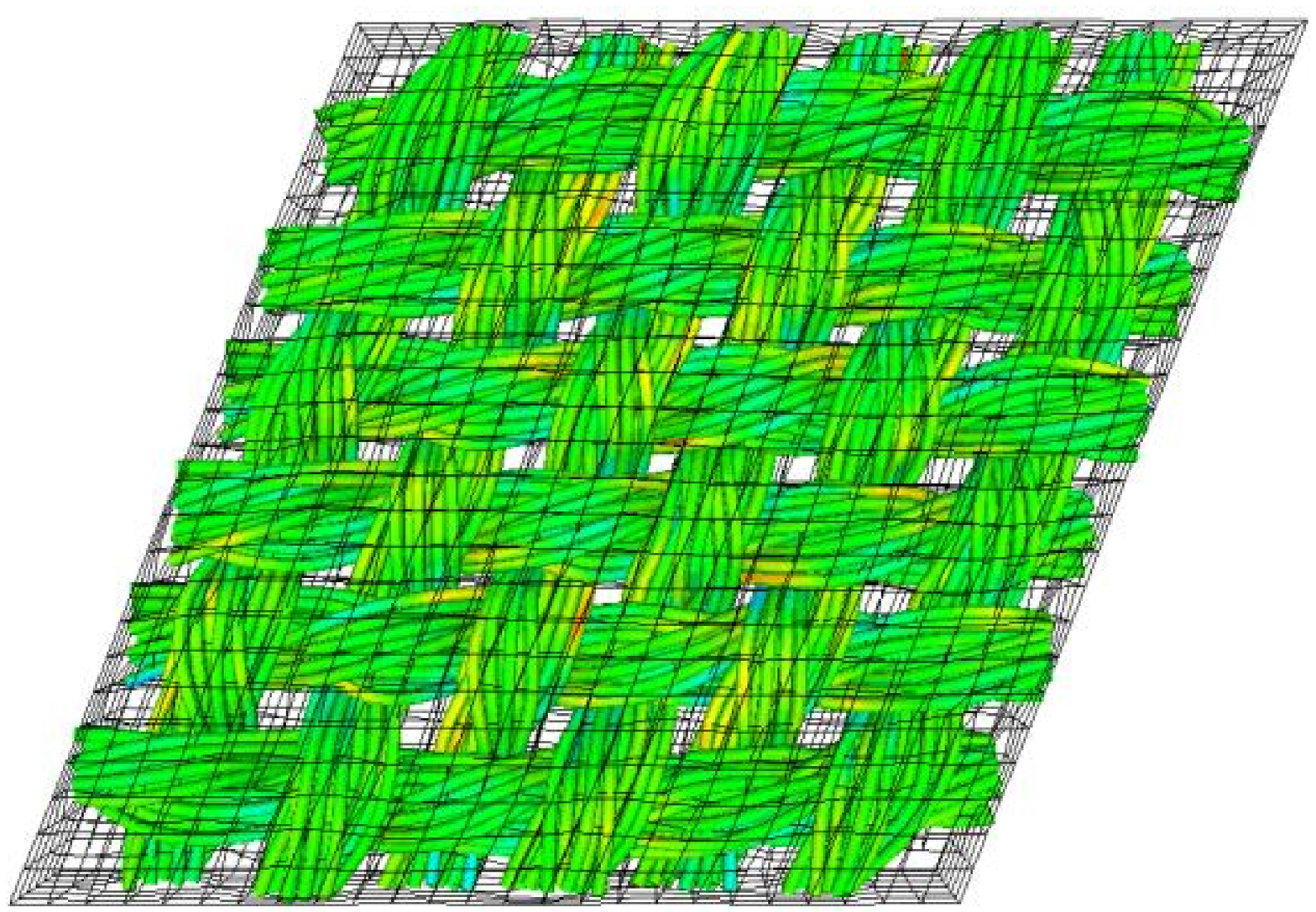} \includegraphics[width=8cm]{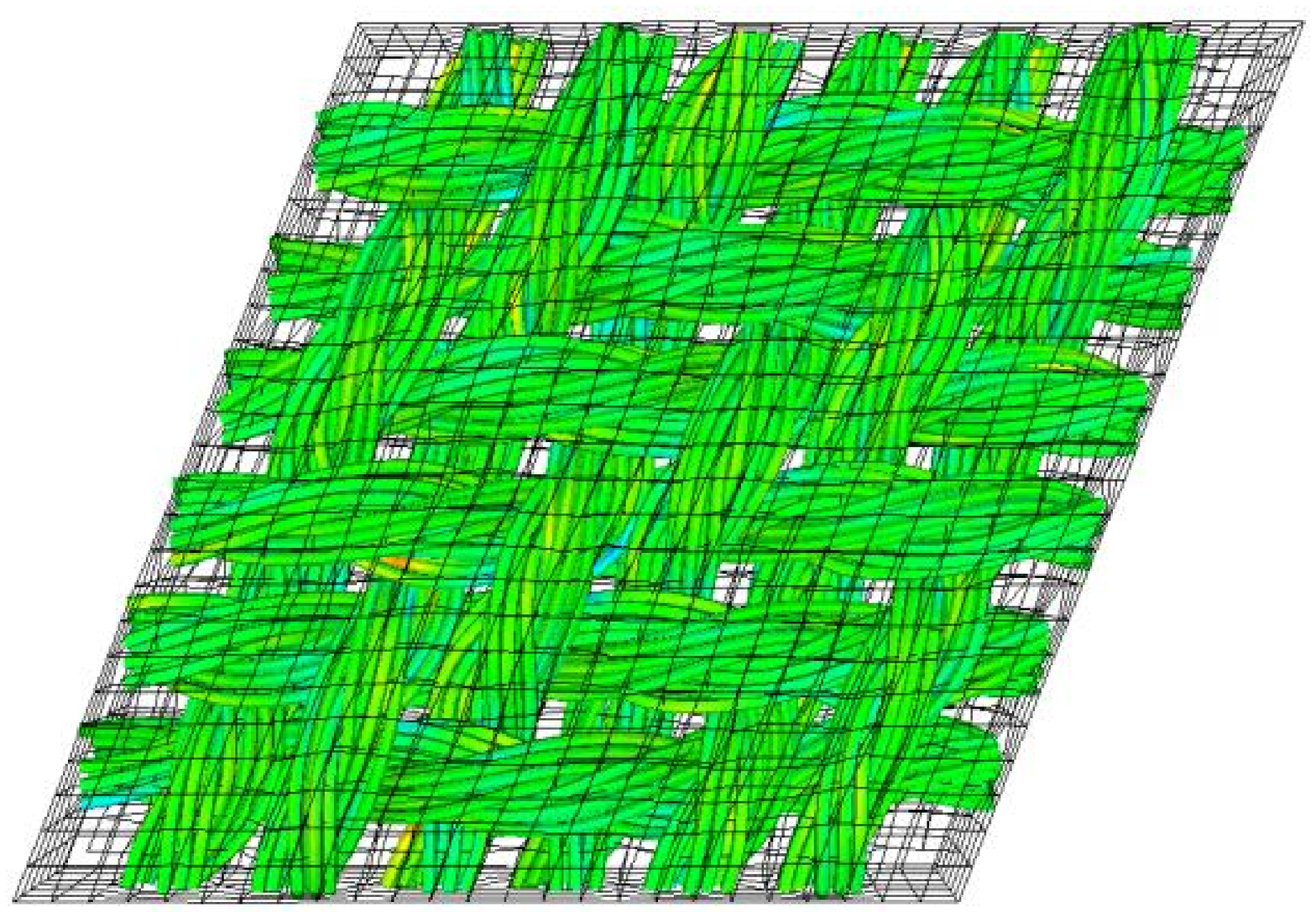}}
  \caption{Deformed configurations at the end of the shear loading}
  \label{fig:ShearDeformMeshes}
\end{figure}

\section{Conclusion}

A global model to simulate and identify the constitutive behaviour of textile composites have been
presented. Because it takes into account all components and mechanisms at the meso scale of this
type of structure, this model is able to represent the complex behaviour of textile structures
without having to fit mechanical parameters. The numerical tests show the efficiency of the
algorithms used in the model, even when a large number of contact elements are considered. The use
of this model should allow to understand and estimate the phenomena taking place at the
meso-scale in textile structures, and to identify the global behaviour of textile materials under a
wide range of loading cases.

\end{document}